\documentclass[preprint]{elsarticle}

\usepackage{graphicx}
\usepackage{dcolumn}
\usepackage{bm}
\usepackage{amssymb}
\usepackage{comment}
\usepackage{hyperref}
\usepackage[utf8]{inputenc}
\usepackage[T1]{fontenc}
\usepackage{lmodern}
\usepackage{arydshln}
\usepackage[ruled,vlined]{algorithm2e}
\usepackage{amsmath}

\usepackage{subfigure}
\usepackage{xcolor}

\begin{document}


\title{Empirical Growing Networks vs Minimal Models: Evidence and Challenges from Software Heritage and APS Citation
Datasets}

\author{Guillaume Rousseau}
\address{Laboratoire Matières et Systèmes Complexes, UMR7057 CNRS and Université Paris Cité,
10 rue Alice Domon et Léonie Duquet, F-75013 Paris cedex 13, France}

\date{\today}

\begin{abstract}
We investigate the evolution rules and degree distribution properties of the
{\it Software Heritage} dataset, a large-scale growing network linking software
source-code versions from open-source communities. The network spans more than
40 years and includes about $6 \times 10^9$ nodes and edges. Our analysis relies
on deterministic temporal and topological partitions of nodes and edges, which account
for the multilayer and partially time-stamped structure of the {\it main graph}.
We derive a {\it temporal graph} that reveals a mesoscale structure and
enables the study of edge dynamics—creation, inheritance, and aging—together with
comparisons to minimal models using 
degree distributions and
histograms of edge timestamp differences. The {\it temporal graph}  also exposes
regime shifts that correlate with changes in developer practices, as reflected
in the average number of edges per new node.
We estimate scaling exponents under the scale-free hypothesis and highlight 
the sensitivity of the estimation method used to both
regime shifts and outliers, while showing that partitioning improves regularity
and helps disentangle these effects. We extend the analysis to the APS citation
network, which also exhibits a major regime shift, with an accelerated growth
regime becoming dominant after 1985. Although both datasets are 
{\it a priori} good candidates for advanced quantitative analysis,
our results illustrate how structural and dynamical transitions hamper our
ability to draw firm conclusions about the existence and observability of a
scale-free regime in these empirical networks. These findings underscore the
need for refined tools and models to study transient growth regimes, to extend
current frameworks toward minimal causal growth models, and to enable robust
comparisons between empirical growing networks and minimal models.
\end{abstract}


\maketitle

\section{Introduction}
Studying the dynamical properties of complex systems through their representation
as networks remains a central approach across physics, biology, chemistry, and
the social sciences.
Network models provide a unified framework for the analysis of real-world systems 
by describing how interacting entities give rise to large-scale structures and emergent behaviors.
Such interaction rules are most often formulated as probabilistic mechanisms acting on nodes and edges, 
and define minimal models that capture a wide range of growth regimes observed in real-world networks
\cite{albert_statistical_2002}.

Since the role of preferential attachment in the emergence of scale-free networks was recognized,
considerable attention has been paid to their topological
properties, and more particularly to degree distributions. The main mechanisms
leading to different asymptotic parametric families of degree distribution
functions (power law, exponential, lognormal, pure or with deviations, such as cutoff,
\ldots) have been identified through theoretical studies in the
large-scale, long-time limit of minimal models.

\subsection{Current challenges}

Twenty years later, despite the existence of this theoretical framework and several corpora
gathering hundreds of real-world network datasets, there is still no agreed-upon,
standardized methodology (nor any comprehensive toolbox) to analyze observed data and
relate them to the taxonomy of networks emerging from minimal models. Several challenges
remain open and contribute to this situation:

{\it -- Methods for measuring and analyzing network properties}, to test the agreement
between hypothesized mechanisms and data. Challenges include the development of robust
techniques~\cite{pham_pafit_2015,arnold_likelihood-based_2021,inoue_joint_2020,
hhk_2020,dimitrova_graphlets_2020,falkenberg_identifying_2020,voitalov_scale-free_2019}
to infer the characteristics of distributions or attachment rules~\cite{sheridan_preferential_2018}, 
while accounting for
finite-size effects and the scale-invariance 
hypothesis~\cite{dorogovtsev_size-dependent_2001,clauset_power-law_2009,serafino_true_2021}, as well as
noise~\cite{kavran_denoising_2021}, outliers, and persistent initial conditions affecting
distribution tails~\cite{dorogovtsev_structure_2000}.

{\it -- Existence of hidden structural building blocks}.
Real-world networks often exhibit mesoscopic structures in which groups of nodes share
common but {\it a priori} unknown properties, giving rise to densely connected clusters or
``communities''. A wide range of community-detection methods has therefore been developed
to capture these structural modules~\cite{fortunato_community_2010,rossetti_community_2018}.
The use of such methods sometimes requires working on networks derived from the original
graph~\cite{palla_uncovering_2005}, which may introduce one or several additional
parameters. More general approaches are not focused on densely connected clusters, but
instead rely on stochastic block models that capture a wider range of mesoscale organization
patterns~\cite{holland_stochastic_1983,peixoto_bayesian_2019,matias_statistical_2017}.

{\it -- Potential changes in evolution rules}. Studies of networks whose evolution spans
decades show changes in associated minimal model parameters, such as an increasing number
of edges per new node~\cite{albert_statistical_2002,meusel_graph_2014}, or at least attempt
to account for this potential issue~\cite{redner_citation_2005,sheridan_preferential_2018}.
More generally, the evolution of the model itself must be considered, including shifts
between different preferential attachment rules~\cite{falkenberg_identifying_2020,bhamidi_change_2018} 
or competition between coexisting
models~\cite{bianconi_competition_2001,arnold_likelihood-based_2021}, which likely result in transient 
phenomena~\cite{barabasi_evolution_2002} and further complicate these studies.

\subsection{Methodological concerns}

When analyzing the properties of real-world networks, existing approaches 
are often complementary; however, they may sometimes raise methodological issues. 
In this work, we focus on two empirical datasets, 
namely the Software Heritage (SWH) and APS citation datasets. 
Both datasets have heterogeneous, multilayer structures in which only some node types carry temporal information.
This directly motivates the methodological considerations discussed below:

{\it -- Scale-free networks and community detection}.
Minimal random models that generate scale-free degree distributions typically produce
graphs with very low clustering coefficients, which decreases with system size
\cite{albert_statistical_2002}. Consequently, scale-free networks are often considered
as null models for community-detection algorithms. Several benchmarking strategies
therefore rely on ad hoc modular networks derived from systems exhibiting
scale-invariant properties. Such test cases are regarded as particularly challenging and
thus relevant for assessing and comparing different community-detection techniques (see
Sec.~XV.A of \cite{fortunato_community_2010}).

{\it -- Missing temporal information}. Whether the focus is on community dynamics,
microscopic evolution rules, or scale-invariant properties, the use of temporal
information is central, as it affects all subsequent analyses and may introduce
significant methodological biases. When temporal information is partially missing,
defining a derived {\it temporal graph} becomes conceptually related to inferring
node-level attributes from structural properties, in a way reminiscent of how nodes may
be assigned to subsets in community-related analyses. In most studies, 
such a procedure should not be interpreted as community or block detection, but rather as a preliminary
step in which partitioning algorithms are commonly employed in the study of real-world
networks, for instance when transforming a weighted directed graph into an undirected
binary one \cite{palla_uncovering_2005}. 
More importantly, properties inferred from such derived graphs must be tested for
robustness with respect to changes in the construction procedure or in its parameters
before being interpreted as intrinsic properties of the system.

{\it -- Very large/old networks vs growth regime shifts}. The evolution rules governing
the subgraph formed by nodes with temporal attributes are particularly favorable, as they
are simpler than those encountered in many real-world systems: nodes do not disappear
once created, edges are directed, and they appear immediately upon the creation of the
source node. Moreover, the size of these two very large datasets is precisely one of the
characteristics sought by researchers, in line with the long-term hypotheses mentioned
above; however, it also increases the probability that changes in the underlying evolution
rules may occur.

{\it -- Multilayer networks and community detection}. This multilayer organization
naturally defines distinct structural components. This does not preclude the existence of
implicit communities or blocks, but any such analysis must take these native layers into
account. In the APS dataset, for example, overlapping scientific communities in a
co-author network \cite{palla_uncovering_2005} coexist with the journal layer available; 
in the SWH dataset, known microscopic rules favor
the formation of triangles and clusters linked to project-level structures, leading to
the empirical classification of fork types \cite{pietri_forking_2020}. As in
community-level evolution studies, partitioning the graph using existing cliques helps
disentangle overlapping substructures. Beyond such local or mesoscopic approaches,
network organization can sometimes be described by {\it hierarchical minimal models},
where hierarchical modularity can be assessed through the dependence of the clustering coefficient 
of a node on its degree (see Chap.~9 of \cite{barabasi_network_nodate}).
This, in turn, raises again the question of a unified methodological framework,
particularly given the complementarity of existing approaches
\cite{boccaletti_structure_2014,fortunato_community_2010}.

\subsection{Problem statement and objectives}

Unless stated otherwise, ``partitioning'' is used in a general sense and refers to the
construction of derived graphs on which evolution rules can be studied. Two partitioning
strategies are introduced: {\it temporal partitioning} propagates timestamps to parts of
the graph lacking native temporal information, producing a derived {\it temporal graph},
while {\it topological partitioning} relies on in- and out-degrees as well as on the
presence or absence of self-loops. 
The resulting description can be microscopic or mesoscopic, without requiring one to specify 
whether it arises from community or block detection techniques or from intrinsic properties 
of heterogeneous multilayer networks.

In this context, the objective of the present work is to investigate, using the
SWH and APS citation datasets, to what extent partitioning of multilayer,
partially timestamped networks enable the construction of derived graphs, the isolation of
growth mechanisms, and their relation to minimal generative models.

\section{Result Summary}

Focusing first on the SWH dataset and its {\it main graph} 
(Sec.~\ref{sec:dataset}), the study of nodes with native
temporal attributes (Sec.~\ref{sec:subgraphSWH}) shows how topological partitioning based
on out-degree distributions reveals changes in evolution rules, coinciding with changes
in software development workflows, namely the adoption of $git$ in developer communities
from around 2010 onward. 

While identifying the transition between growth regimes is
straightforward after topological partitioning, out-degree and in-degree
distributions exhibit a highly irregular pattern, partly due to ``outlier'' events
that impact the shape of the observed distributions.

Subgraphs of specific software development projects (e.g., Linux, PHP Composer)
provide evidence for competing growth mechanisms and suggest that a complete
description of the evolution rules must also account for nodes without native temporal
attributes. We therefore derive several graphs from the {\it main graph} by propagating
temporal information up to origin nodes and by applying parametric $TSL$ topological
partitioning (Sec.~\ref{sec:derivedgraphSWH}).

These derived graphs make it possible to study the evolution rules governing the top layer
of the {\it main graph}, providing a first insight into the global dynamics of the SWH
network at the scale of open-source communities and allowing us to discuss growth
mechanisms such as edge creation, inheritance, and aging in a way that can be directly
compared with minimal models.
The comparison (Sec.~\ref{sec:comparison}) is then mainly based on the analysis of the
temporal evolution of in- and out-degree distributions, histograms of edge timestamp
differences, and estimates—under the scale-free hypothesis—of the scaling exponents
associated with the tail of the in-degree distributions.

We illustrate this approach by applying it to another empirical system, 
namely the APS citation network (Sec.~\ref{sec:application}),
and briefly discuss (Sec.~\ref{sec:discussion}) the generality of our findings and how this
study may contribute to the development of a generic methodology for analyzing real-world
growing networks and comparing them with minimal models. 

The analysis shows that, contrary to common assumptions, the APS dataset also exhibits a
significant change in its evolution rules over the time span covered by the data, 
like what is observed in the SWH dataset. However, in contrast to the SWH dataset, whose
final regime appears compatible with the hypothesis of a constant average number of new
edges per new node after the transition, the APS dataset exhibits a transition toward an
accelerated growth regime, which becomes dominant at least from 1985 onward.

All supplemental materials, including the Python scripts required to reproduce the study
from the publicly available raw dataset, are available on a GitHub repository~\footnote{\url{https://github.com/grouss/growing-network-study}}.

\section{SWH Dataset Description}
\label{sec:dataset}
\begin{figure}
    {\centering
    \hbox{
  \includegraphics[width=1.0\linewidth]{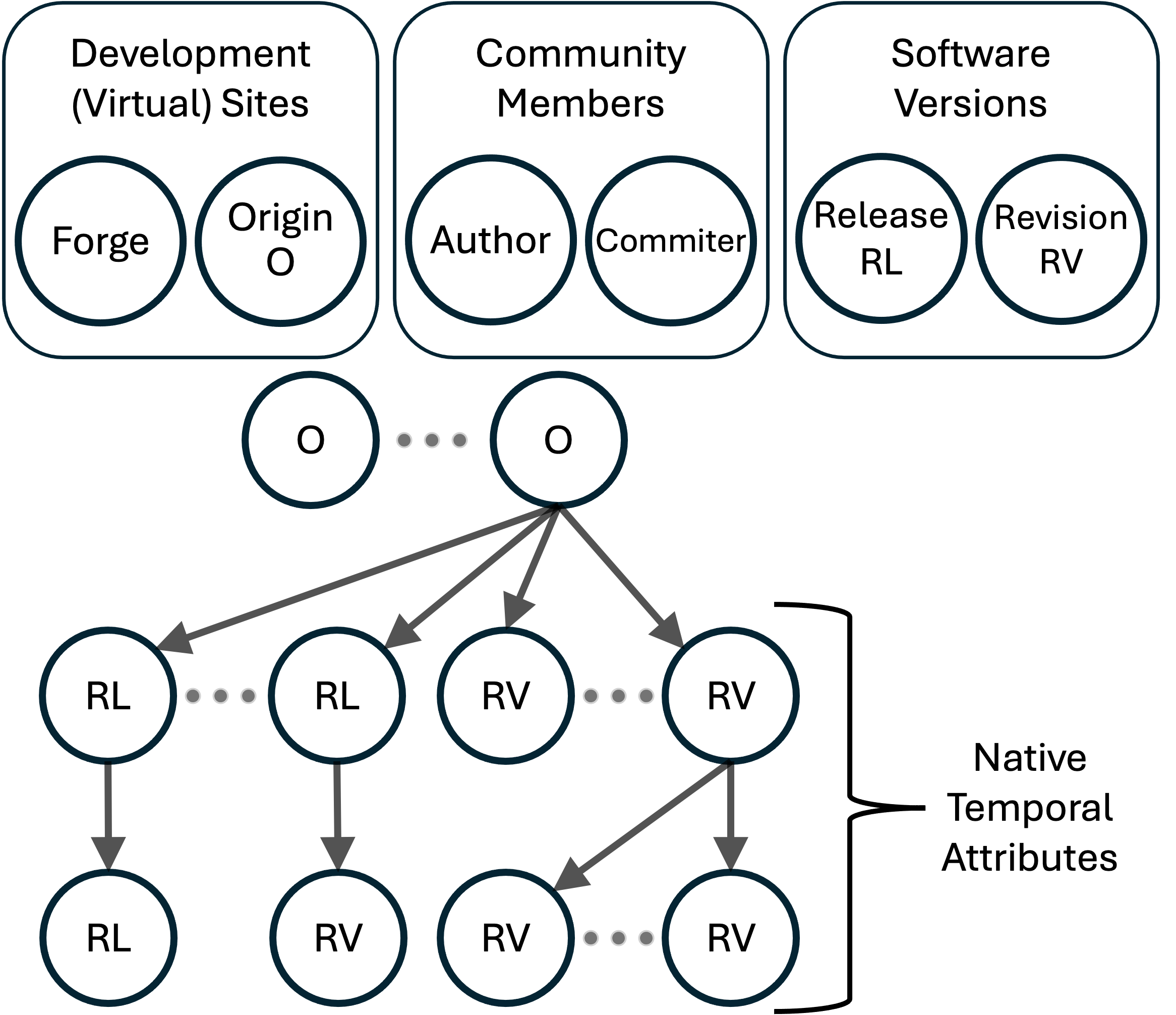}
    }
    }
    \caption{
    Graph representation of the SWH dataset studied here (the {\it main graph}), where nodes
represent software versions ({\it releases}/{\it revisions}) and artifacts produced by
projects across various {\it origins}/{\it forges}. Developers can act as {\it authors}
and/or {\it committers} within these projects. Release and revision nodes include native
temporal attributes linked to committer or author dates. Edge directions follow multilayer
rules and may depend on the nodes' intrinsic identifiers. The lower layers associated with
$RV$ and $RL$ nodes form a directed acyclic graph (DAG).
}
  \label{fig:dataset}
\end{figure}

We now briefly describe the SWH dataset used in this study. Figure~\ref{fig:dataset}
represents a simplified version of the graph extracted from the Software Heritage project
\cite{abramatic_building_2018}, which collects software source code from open-source
communities. For our analysis, we first consider software versions,
understood as snapshots of the source code at a given time. These include two subtypes:
$RV$, representing {\it revisions}, and $RL$, representing {\it releases}, each uniquely
identified by an intrinsic identifier. The directed edges between $RV$ and $RL$ nodes
represent ancestor/descendant relationships, tracking the previous version(s) from which
each version derives. Other nodes represent open-source software development {\it origins}
$O$, i.e., individual projects hosted on public {\it forges} (e.g., \textit{github.com},
\textit{gitlab.com}). These platforms typically aggregate large numbers of such origins. Edges
between origin nodes and $RV/RL$ nodes represent the locations from which the
corresponding software versions were extracted by the Software Heritage crawler.

The results presented here are based on a snapshot of the SWH dataset
from March~23, 2021~\cite{swhaws}, which we refer to as the {\it initial dataset}.
It includes nearly $10^{10}$ nodes, including approximately $2\times 10^9$ software
releases and revisions and around $1.3\times 10^8$ origins (see the Replication Package
for details).

In this network, temporal information is found in software versions ($RV$ and $RL$ nodes)
through one timestamp corresponding to the {\it commit} date of the version (i.e., the
date when it was made available to other developers of the project via the source code
version management tool), and possibly a second timestamp associated with the author's
commit date, if the author differs from the committer. Because nodes are identified by
intrinsic identifiers associated with software versions, the lower layers form a directed
acyclic graph (DAG). For more details, we refer to previous studies of this dataset
\cite{rousseau_ese,pietri_determining_2025}, the latter including a discussion of suitable
graph representations for analyzing intrinsic properties at scale.

\begin{figure}
    {\centering
    \hbox{
  \includegraphics[width=1.0\linewidth]{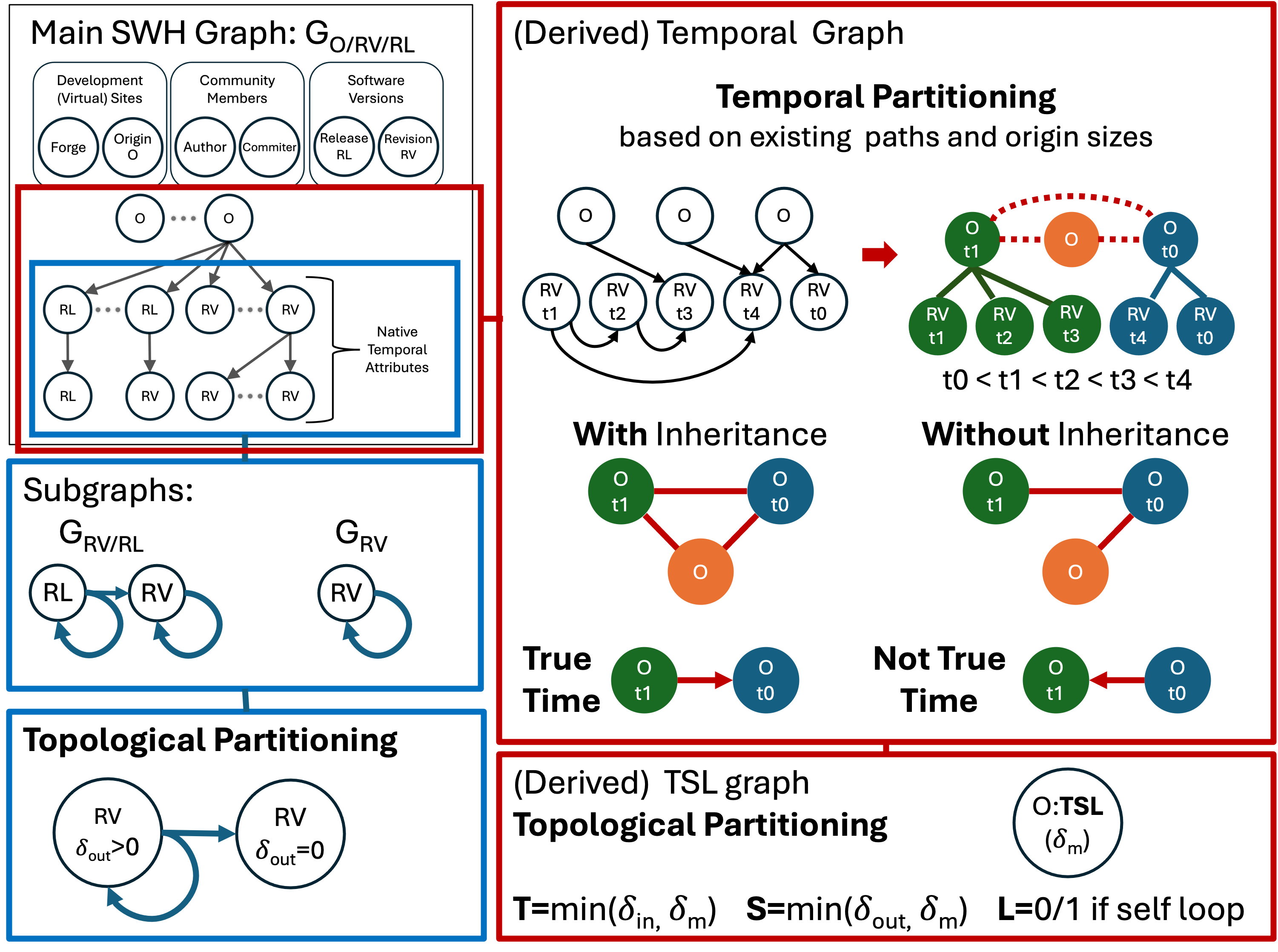}
    }
    }
    \caption{
Overview of the graph processing pipeline used in this work. Starting from the SWH
{\it main graph}, we extract several subgraphs (blue frames) corresponding to nodes with
native temporal attributes and define a derived {\it temporal graph} (red frames) by
partitioning $RV$ and $RL$ nodes according to existing paths and origin sizes, and by
propagating their temporal information to the corresponding origin nodes. Two parameters
allow us to build variants of the {\it temporal graph} using different inheritance and
edge-orientation rules. The resulting {\it temporal graph} is then transformed into a
{\it TSL graph} through topological partitioning.
}
  \label{fig:partitioning}
\end{figure}

Initially, we focus on the subgraph of nodes with native temporal attributes
(Sec.~\ref{sec:subgraphSWH}) and then extend the analysis to the entire graph by
constructing two derived graphs, namely the {\it temporal graph} and the {\it TSL graph},
using temporal and topological partitionings (Sec.~\ref{sec:derivedgraphSWH}). The
notations corresponding to the different graphs, nodes, and edges are summarized in
Table~\ref{tab:notations}. Figure~\ref{fig:partitioning} illustrates the overall pipeline.
Each transformation step is described in the corresponding section.

\renewcommand{\arraystretch}{1.3}
\begin{table}[h!]
\centering
\begin{tabular}{m{0.2\textwidth}m{0.8\textwidth}}
\hline
\textbf{Object} & \textbf{Notation: Description} \\
\hline

{\bf Main graph} & $\mathbf{G_{O/RV/RL}=\{O/RV/RL\}{-}\{O/RV/RL\}}$: 
Software Heritage graph studied here, including 
node types $RV$, $RL$, and $O$ (revisions, releases, and origins),
and the directed edges between them ($O{\to}RL$, $O{\to}RV$, $RL{\to}RL$, $RL{\to}RV$, and $RV{\to}RV$).\\\hdashline

Subgraph  &$\mathbf{G_{RV/RL}=\{RV/RL\}{-}\{RV/RL\}}$:
Restricted to release $RL$ and revision $RV$ nodes carrying 
native temporal attributes, with edges between them ($RV{\to}RV$, $RL{\to}RV$ and $RL{\to}RL$).\\\hdashline

Subgraph  &$\mathbf{G_{RV}=RV{-}RV}$:
Restricted to revision nodes $RV$, with edges between them $RV{\to}RV$.\\

\hline

{\bf Temporal graph} & $\mathbf{G^{modeI,modeT}=O{-}(RV/RL){-}O}$: 
Derived graph obtained by propagating the temporal information to the $O$ nodes of the {\it main graph} 
and aggregating directed paths through $RV/RL$ nodes to build $O{\to}O$ edges.
Only contains $O$ nodes, each carrying a temporal attribute $\hat{t}$.
Four variants are defined depending on the inheritance ($modeI\in\{I,WI\}$) 
and true-time ($modeT\in\{TT,NoTT\}$) rules used.
\\
\hdashline

{Inheritance rules} &
$\mathbf{G^{\mathrm{I,.}}}$, $\mathbf{G^{\mathrm{WI,.}}}$: 
With inheritance, an edge $o_i \to o_j$ is added whenever a directed path exists 
from $o_i$ to $o_j$ through $RV/RL$ nodes after partitioning. Without inheritance, 
an edge $o_i \to o_j$ is added only if there exists at least one edge of
type $RL{\to}RL$, $RL{\to}RV$, or $RV{\to}RV$ connecting a node assigned to $o_i$ 
to a node assigned to $o_j$.\\ \hdashline

{True-time rules} &
$\mathbf{G^{\mathrm{.,TT}}}$, $\mathbf{G^{\mathrm{.,NoTT}}}$: In the true-time variants (TT), 
the direction of an edge $o_i \to o_j$ follows the
true temporal ordering of origins, i.e. $\hat{t}(o_i) > \hat{t}(o_j)$.
In the variants not based on true-time (NoTT), edge directions follow the direction of
the paths through $RV/RL$ nodes assigned to $o_i$ and $o_j$.\\

\hline
{\bf TSL graph} & $\mathbf{G^{TSL}_{\delta_m}}$: Derived from the {\it temporal graph}. 
Edges between $O$ nodes classified according to parametric $TSL$ types (e.g. $111{\to}111$, $111{\to}101$,
$011{\to}101$).\\\hdashline

$TSL$ partitioning & $\mathbf{O:TSL(\delta_m)}$: 
Each origin node of the derived {\it temporal graph} is assigned 
a triplet $(T,S,L)$ where: 
$T = \min(\text{in-degree}, \delta_m)$, 
$S = \min(\text{out-degree}, \delta_m)$, 
$L = 1$ if a self-loop exists, $0$ otherwise. 
For $\delta_m = 1$, this classification yields four main types: $001$, $011$, $101$, and $111$.\\

\hline
{\bf Modified BA (Price) model} & $G_{\mathrm{BA}}$:
Modified Barabási--Albert minimal model with oriented edges, like the Price model.\\

\hline
\end{tabular}
\caption{Summary of the graph definitions and notations used for the SWH dataset. The graph pipeline is represented in
Fig.~\ref{fig:partitioning}. Algorithms are detailed in Sec.~\ref{sec:algorithms}.}
\label{tab:notations}
\end{table}

\section{Growth of nodes and edges over time in the {\it main graph}}
\label{sec:subgraphSWH}
In most minimal models, an implicit timestamp can be derived from the order in which nodes are
added. However, this construction does not generally coincide with the native temporal
scale of node or edge attributes in empirical datasets, and is not necessarily the most
relevant timescale for studying real-world network evolution. This issue is particularly
acute in heterogeneous networks arising from human activity, such as those studied here,
where different types of nodes may follow distinct growth rules and evolve on markedly
different timescales.

\begin{figure}
    {\centering
    \hbox{
  \includegraphics[width=1.0\linewidth]{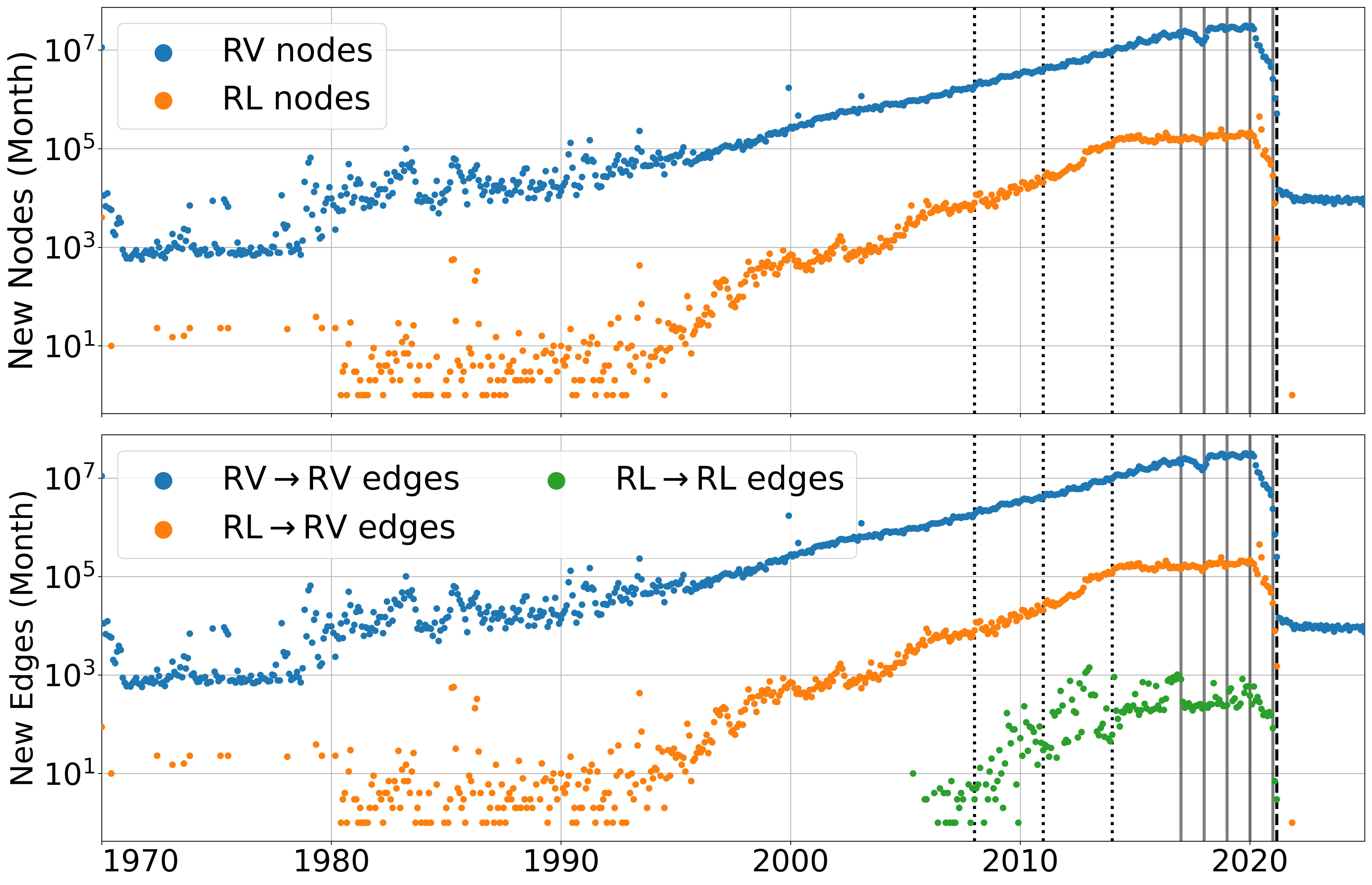}
    }
    }
    \caption{
    New nodes (top) and edges (bottom) per month by type ($RV$: revision, $RL$: release) from
1970 to 2030 in the {\it main graph} of SWH dataset (exported March 2021, dashed line). 
Exponential growth is
observed, except for $RL$ nodes and the associated $RL{\to}RL$ and $RL{\to}RV$ edges,
which exhibit a constant rate since early 2014 (third dotted line). The appearance of
$RL{\to}RL$ edges aligns with the adoption of $git$ and the launch of $github.com$ in 2008
(first dotted line). Plain vertical lines indicate January~1st of each year from 2017 to
2021. Anomalies at the end of 2017 and 15 months before export suggest biases due to SWH
crawling policies. Post-export nodes highlight temporal data issues (see Supplemental
Material).
    }
    \label{fig:edgesnodes}
\end{figure}

In the SWH dataset, only $RV$ and $RL$ nodes have native temporal attributes, 
spanning more than 50 years.  We start with a first, straightforward
analysis of the subgraph induced by the set of nodes with temporal attributes. We consider
the monthly average number of new $RV$ and $RL$ nodes and compare them to the
monthly average number of new edges, taking into account the three possible edge types:
between two $RV$ nodes ($RV{\to}RV$), between two $RL$ nodes ($RL{\to}RL$), or from an
$RL$ node to an $RV$ node ($RL{\to}RV$)\footnote{We use the notation $RV{\to}RV$ to denote
an edge directed from one RV-type node to another, to emphasize both the directed
nature of the edge and to distinguish it from the notation $RV-RV$, which we reserve
to refer to the $RV$-node subgraph in the following (see Table~\ref{tab:notations}).}. 
Figure~\ref{fig:edgesnodes} shows that the numbers of new nodes and edges
added each month follow distinct patterns, with a constant rate of new $RL$ nodes per
month starting in early 2014, while the number of new $RV$ nodes and $RV{\to}RV$ edges
continues to increase exponentially beyond this date.

\begin{figure}
    {\centering
   \hbox{
  \includegraphics[width=1.0\linewidth]{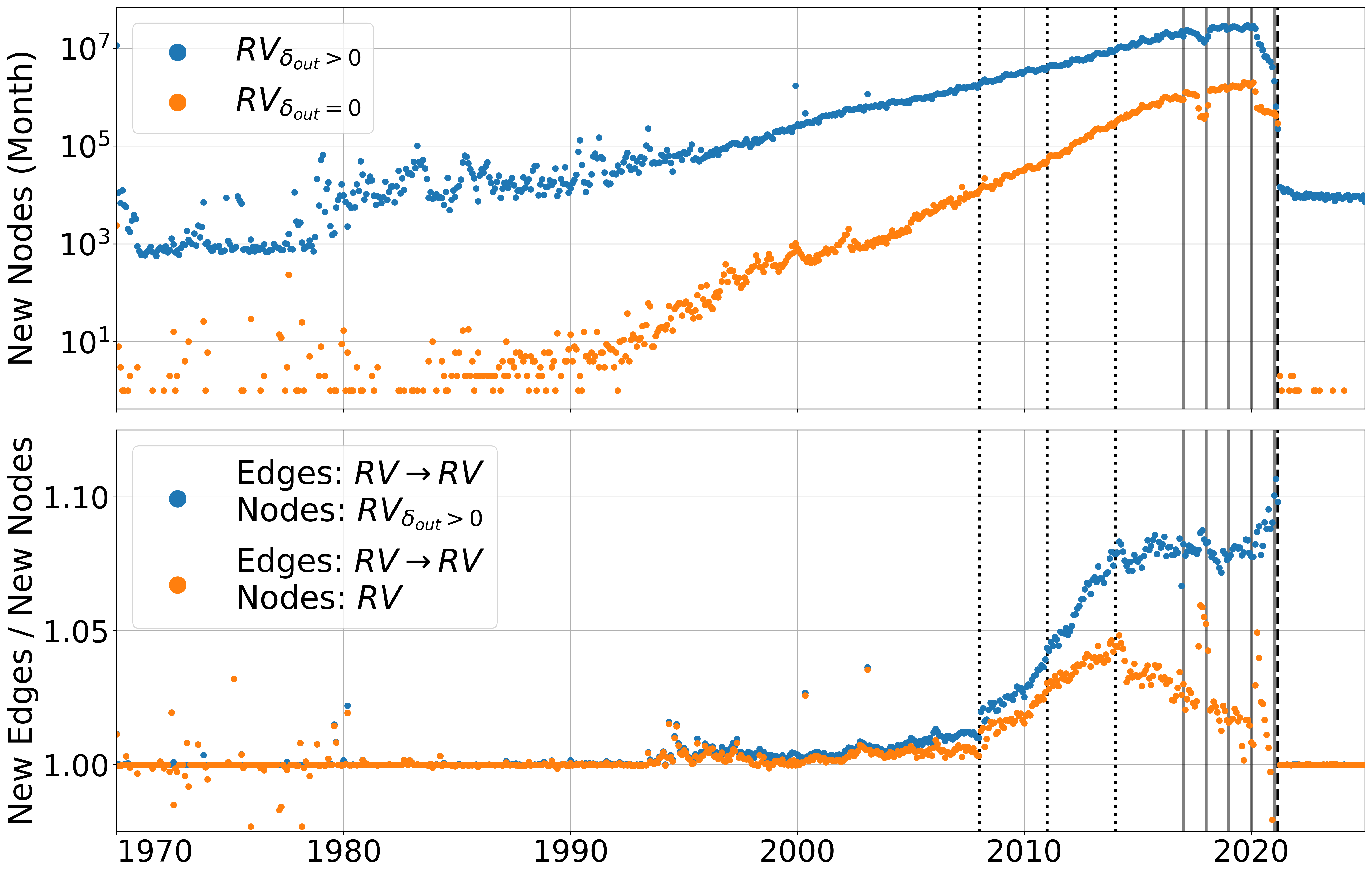}
    }
    }
    \caption{
(Top) Number of new $RV$ nodes and $RV{\to}RV$ edges per month, distinguishing nodes with
outgoing edges ($\delta_{out}>0$) and without ($\delta_{out}=0$). (Bottom) Rate comparison
of new edges per new $RV$ node when considering all nodes (orange) and when restricting
to nodes with $\delta_{out}>0$ (blue). This partitioning reveals an exponential growth
from the mid-2000s to 2013, followed by a constant rate after 2014. In the bottom panel of
Fig.~\ref{fig:edgesnodes}, this rate matches that of new $RL{\to}RV$ edges (orange dots)
but not that of $RV{\to}RV$ edges (blue dots). The post-2014 decrease in the
$RV{\to}RV/RV$ rate reflects the faster growth of $RV$ nodes without outgoing edges
($\delta_{out}=0$) compared to those with at least one outgoing edge.
    }
  \label{fig:ratioRVRV}
\end{figure}

Before proceeding with any comparison to minimal models, it is necessary to determine
whether the exponential growth observed in the number of $RV$ nodes is representative of
those participating in the formation of edges within the network---namely, nodes
with non-zero degree. This corresponds to a topological partitioning which, in practice,
consists in creating two subcategories of $RV$ nodes, denoted $RV_{\delta_{out}=0}$ and
$RV_{\delta_{out}>0}$, associated with $RV$ nodes whose out-degree is equal to zero or
greater than zero, respectively. If relevant, one could further distinguish the associated
edge types, but since the source node of an edge is, by definition, such that
$\delta_{out}>0$, we focus on the monthly averages and ratios involving $RV{\to}RV$,
$RV_{\delta_{out}>0}$, and $RV_{\delta_{out}=0}$. Fig.~\ref{fig:ratioRVRV} shows 
distinct growth regimes, the last of which
is consistent with the minimal model's assumption of a constant number of new
$RV{\to}RV$ edges per new $RV_{\delta_{out}>0}$ node starting from early 2014, and is
similar to what is observed for $RL$ nodes and $RL{\to}RV$ edges
(Fig.~\ref{fig:edgesnodes}).


It follows that the {\it main graph} restricted to $RV$ nodes
and the edges between them (i.e. the {\it $RV-RV$ subgraph}), 
is particularly relevant for a more detailed comparison with
minimal models, which is carried out in Sec.~\ref{sec:comparison}.

\section{Building derived temporal graphs}
\label{sec:derivedgraphSWH}
Some mechanisms may depend on the nature of the project, making it necessary not to limit
the study to the layer including $RV$ and $RL$ nodes, but also to consider $O$ nodes 
to analyze the overall network dynamics. A case-by-case analysis of $RV$ nodes with
the highest number of incoming edges suggests the existence of at least two distinct
growth mechanisms, referred to as ``internal'' and ``external'' (see Supplemental
Material~09).  The {\it internal} mechanism is related to the use of distributed version 
control tools and depends on the size of the teams involved in software development activities, 
as well as on the maturity of the projects.  The {\it external} mechanism corresponds to the
creation of new origin nodes akin to ``forks'' of the project, where the goal, for at
least some of them, is not to create a new project but rather a personalized version of
the original one.

\subsection{Temporal partitioning}

This raises the more general question of how to propagate temporal information to nodes
that do not natively carry it in heterogeneous multilayer networks. For nodes in a DAG
downstream of nodes with native timestamps, a temporal election principle can be
applied~\cite{rousseau_computer-based_2013,rousseau_computer_2011}. For upstream nodes, such as $O$
nodes in the {\it main graph}, temporal partitioning makes it possible to construct a
derived graph in which timestamps are assigned based on those of the partitioned
downstream nodes.

We then introduce a derived {\it temporal graph} (Fig.~\ref{fig:partitioning}) by
propagating temporal information to all $O$ nodes in the {\it main graph} and by defining
aggregated links between $O$ nodes according to the existing directed paths, following
the steps below:
\begin{itemize}
\item For each $O$ node, we define its origin size as the number of downstream reachable
nodes (i.e., $RV$ and $RL$ nodes).
\item Each $RV/RL$ node is assigned to a unique $O$ node, by default the $O$ node with the
largest origin size among all $O$ nodes from which the $RV/RL$ node is reachable.
\item A temporal attribute is assigned to each $O$ node, by default the oldest timestamp
among all $RV/RL$ nodes assigned to that $O$ node.
\item For each $O_i$ node, we build the list of reachable $O_j$ origins, by default (i.e.,
{\it with inheritance rule}) defined as the list of $O$ nodes to which the $RV/RL$ nodes
reachable from $O_i$ have been assigned.
\item Alternatively, if the construction rule is set to {\it without inheritance}, edges
are not created between origins that are reachable only through $RV/RL$ nodes assigned to
a third origin.
\item Edge directions are set either according to the path direction through $RV/RL$ nodes
({\it not-true-time}) or using the natural time arrow ({\it true-time}), i.e., from the
origin node with the youngest timestamp to the origin node with the oldest timestamp.
\end{itemize}

The detailed algorithmic implementation of this construction is provided in the Appendix
(Sec.~\ref{sec:algorithms}). The discussion of inheritance and true-time rules is deferred
to Sec.~\ref{sec:comparison}.
When applied to the {\it main graph} using the {\it with inheritance} rule, this procedure
yields a derived network linking the $O$ nodes together. It
contains 139,524,533 nodes and 80,734,013 edges (see Supplemental Material~10).

\subsection{Topological partitioning}

To generalize the topological partitioning introduced in the study of the $RV$--$RV$
subgraph, we define a classification based on the topological properties of the derived
{\it temporal graph}, which yields the {\it TSL graph} (Fig.~\ref{fig:partitioning} and
Table~\ref{tab:notations}). In this classification, each node is characterized by the
in-degree $T$, the out-degree $S$, and a boolean $L$, which equals $1$ if it links to
itself and $0$ otherwise. Self-loops exist for origin nodes that have one or more $RV/RL$
nodes after partitioning. To limit the number of distinct categories (which may correspond
to different evolution rules), we also introduce the {\it classification depth} $\delta_m$,
which corresponds to the maximum value of $T$ and $S$ used to define categories and
partition the origin nodes. Each origin in the {\it TSL graph} is then assigned a type,
denoted $O:TSL(\delta_m)$ (or simply $TSL$ when not ambiguous), corresponding to the values
$\min(T,\delta_m)$, $\min(S,\delta_m)$, and $L$.

Cycles do not exist in the {\it main graph}, which is a directed acyclic graph, but may
exist in the {\it temporal graph}, depending on the partitioning strategy and on whether
or not the time arrow is used to define edge directions between nodes with one or more
incoming and outgoing edges. For $\delta_m = 1$, cycles can only involve $TSL$ 
nodes classified as $111$ after topological partitioning.

\begin{figure}
    {\centering
    \hbox{
    \includegraphics[width=1.0\linewidth]{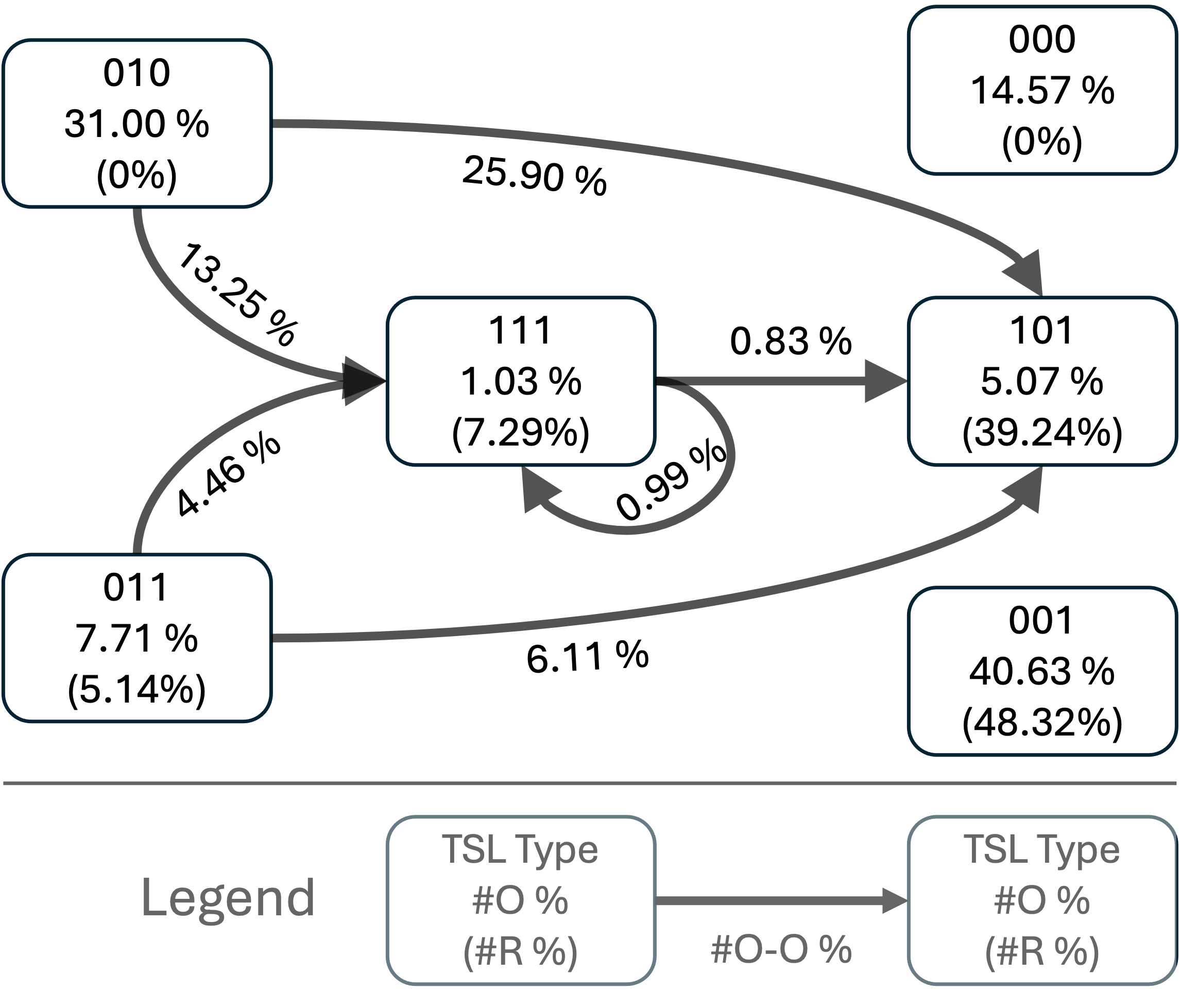}
    }
    }
    \caption{
Adjacency diagram of the {\it TSL graph}. This representation shows the weights of the
different $TSL$-type origin nodes ($\delta_m = 1$). 
Self-loops are included in the edge-weight normalization, 
which explains why the sum is smaller than 100\%.
Percentages in parentheses correspond to the ratio of $RV$ and $RL$ nodes assigned after
partitioning by $TSL$ type. Origin nodes of type $111$ and $101$ account for only a small
fraction of all origin nodes (1\% and 5\%, respectively), despite playing a central role in
the network's growth. In contrast, nodes of type $001$, which represent approximately
40\% of all origin nodes and 48\% of $RV/RL$ nodes, act primarily as reservoir nodes for
$101$ and $111$ nodes, which together account for about 6\% of origin nodes and 46\% of
$RV/RL$ nodes.
    }
    \label{fig:TSLgraph}
\end{figure}

From this perspective, starting from the {\it main graph}, in which only some nodes
natively carry a temporal attribute and whose corresponding layer contains no cycles, the
construction of the derived {\it temporal} and {\it TSL graphs} makes it possible to
represent the upper layer of the system as a directed graph in which cycles may exist. One can then introduce
the adjacency diagram describing the links between the different $TSL$ types
(Fig.~\ref{fig:TSLgraph}). 

Previous studies of the Web graph have shown how such
adjacency diagrams are organized into strongly connected components, together with
subsets of nodes that do not belong to these components but for which incoming or outgoing
paths may exist. This has led to the seminal results describing the bow-tie structure of
the Web~\cite{broder_graph_2000}, and to subsequent works~\cite{meusel_graph_2014,meusel_graph_2015}.

In these representations, the ``OUT'' class corresponds to nodes whose edges are all
incoming (i.e., $T \ge 1$ and $S = 0$), whereas the ``IN'' class corresponds to nodes whose
edges are all outgoing (i.e., $T = 0$ and $S \ge 1$). A detailed discussion of these
structures, and of their relation to the distribution of strongly connected
component sizes, is beyond the scope of the present study, although some elements are
already provided in the Supplementary Materials. It nevertheless follows that, thanks to
the temporal and topological partitionings, 
it becomes possible to study both the mesoscopic structure of the
system and its evolution rules at the system scale.

The $TSL$ adjacency diagram and a systematic study of the evolution rules based on the
$TSL$ types show that the degree distributions of the {\it temporal graph} are dominated by
$011{\to}111$ and $011{\to}101$ edges, thereby masking part of the underlying growth
mechanisms (see Supplementary Material~11).
The transition observed in both the {\it main graph} and the {\it temporal graph} around
2009--2011 (see Fig.~\ref{fig:ratioRVRV} and Fig.~\ref{fig:DegreeOOmerged_bottom},
discussed later) should, first and foremost, be interpreted as a transient phenomenon
following the emergence of new types $011$ and $111$. These are directly associated with a
change in practices within the real-world system, namely the adoption of \textit{git} in
developer communities, and can subsequently be interpreted in terms of ``microscopic''
growth rules and $TSL$ types.

\section{Comparison with minimal models}
\label{sec:comparison}
We now assess the relevance of these partitioning strategies for comparisons with minimal
models. We begin by analyzing their impact on the study of in- and out-degree
distributions (Sec.~\ref{subsec:degree_distributions}), which are directly tied to the
construction rules of minimal models. We then turn to another aggregated observable: the
histograms of signed edge timestamp differences,
${\rm sgn}(\Delta TS)\log_{10}(|\Delta TS|)$ (Sec.~\ref{subsec:histogram_deltat}).

We compare the observed quantities with those generated by a modified Barabási--Albert
model with oriented edges, which is similar to the Price
model~\cite{price_general_1976}. This model is hereafter referred to as a
``modified Barabási--Albert (Price) model'', underlining the methodological relevance of
adapting minimal models for comparison with empirical datasets. We fix $m = 2$, the
number of new edges per added node; edges are oriented according to the order of node
appearance, and timestamps are defined to mimic the exponential growth in the number
of nodes observed in the {\it main graph}. The preferential attachment rule considers, 
for each node, the sum of its out-degree (which is fixed and equal to $m$) and its
in-degree. The network is initialized with a complete graph of $m+1$ nodes.

\subsection{In- and out-degree distributions over time}
\label{subsec:degree_distributions}

\begin{figure}
    {\centering
    \hbox{
    \includegraphics[width=1.0\linewidth]{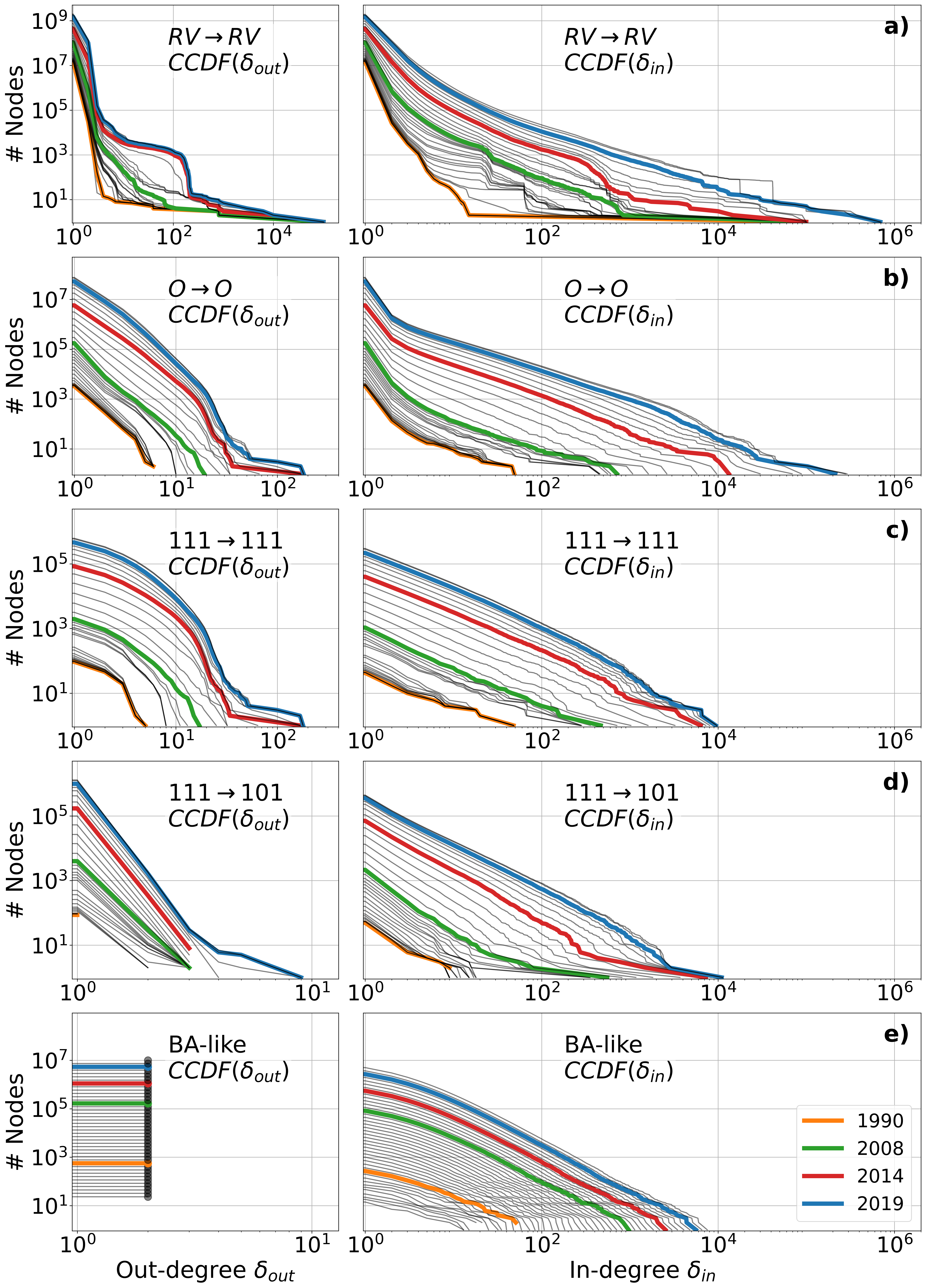}
    }
    }
    \caption{
    Complementary cumulative distribution functions (CCDFs) of out-degrees (left) and
    in-degrees (right) over time. From top to bottom, the panels correspond to:
    $RV{\to}RV$ edges of the {\it main graph}; $O{\to}O$ edges of the {\it temporal graph};
    $111{\to}111$ and $111{\to}101$ edges from the {\it TSL graph} after $TSL$ partitioning;
    and, for comparison, a modified Barabási--Albert (Price) model (with two outgoing edges
    per new node, edges oriented according to node appearance order, and timestamps defined
    to mimic the exponential growth of new nodes observed in the {\it main graph}). The
    distributions are shown for January~1st of 2008, 2014, and 2019, using different colors.
    }
    \label{fig:DegreeOOmerged_top}
\end{figure}

Fig.~\ref{fig:DegreeOOmerged_top} shows the in- and out-degree distributions between 1980
and 2021 for the {\it main graph}, the {\it temporal graph}, two of the $TSL$ partitioning
types, and the distributions obtained from the modified Barabási--Albert (Price) model.
The distributions associated with the {\it temporal graph} (second panel, $O{\to}O$)
appear more regular, less affected by large short-term fluctuations. For instance, the
sharp excess observed in 2014 in both the in- and out-degree distributions associated with
$RV$ nodes in the $RV-RV$ subgraph (top panel) is considerably attenuated.

These local fluctuations can be distinguished from those observed in growth models based 
on preferential attachment rules. 
If they resulted in an increased probability of becoming the target node of
subsequent edges, one would instead expect a shift of these excesses toward higher degree
values, as observed, for instance, in the modified Barabási--Albert (Price) model panel,
whose imprint of fluctuations propagates and persists in the tail of the distribution.

Another characteristic of evolution rules in minimal models is the simplicity of 
their formulation regarding the number of outgoing edges from newly added nodes.
In the case of
the modified Barabási--Albert (Price) model, this number is fixed. This is visible
in the bottom panel, where all nodes in the network have exactly $m = 2$ outgoing edges. In
contrast, several real-world networks, such as the graph of the Web, are known to exhibit
non-trivial out-degree distributions. The distributions shown in the first two panels
({\it main graph} and {\it temporal graph}) may suggest a similar situation. However, the
following two panels (the third and fourth from the top) reveal that the $TSL$ partitioning
of the {\it temporal graph} highlights distinct structural rules---particularly for the
$111{\to}101$ edges.

The $111$ nodes appear to be the source of only a single outgoing edge targeting a $101$
node, with few exceptions, while exhibiting a non-trivial in-degree distribution. This
behavior brings the analysis of this real-world network close to the
characteristics observed in networks generated by minimal models.

\subsection{Histograms of the edge timestamp differences}
\label{subsec:histogram_deltat}

\begin{figure}
    {\centering
    \hbox{
    \includegraphics[width=1.0\linewidth]{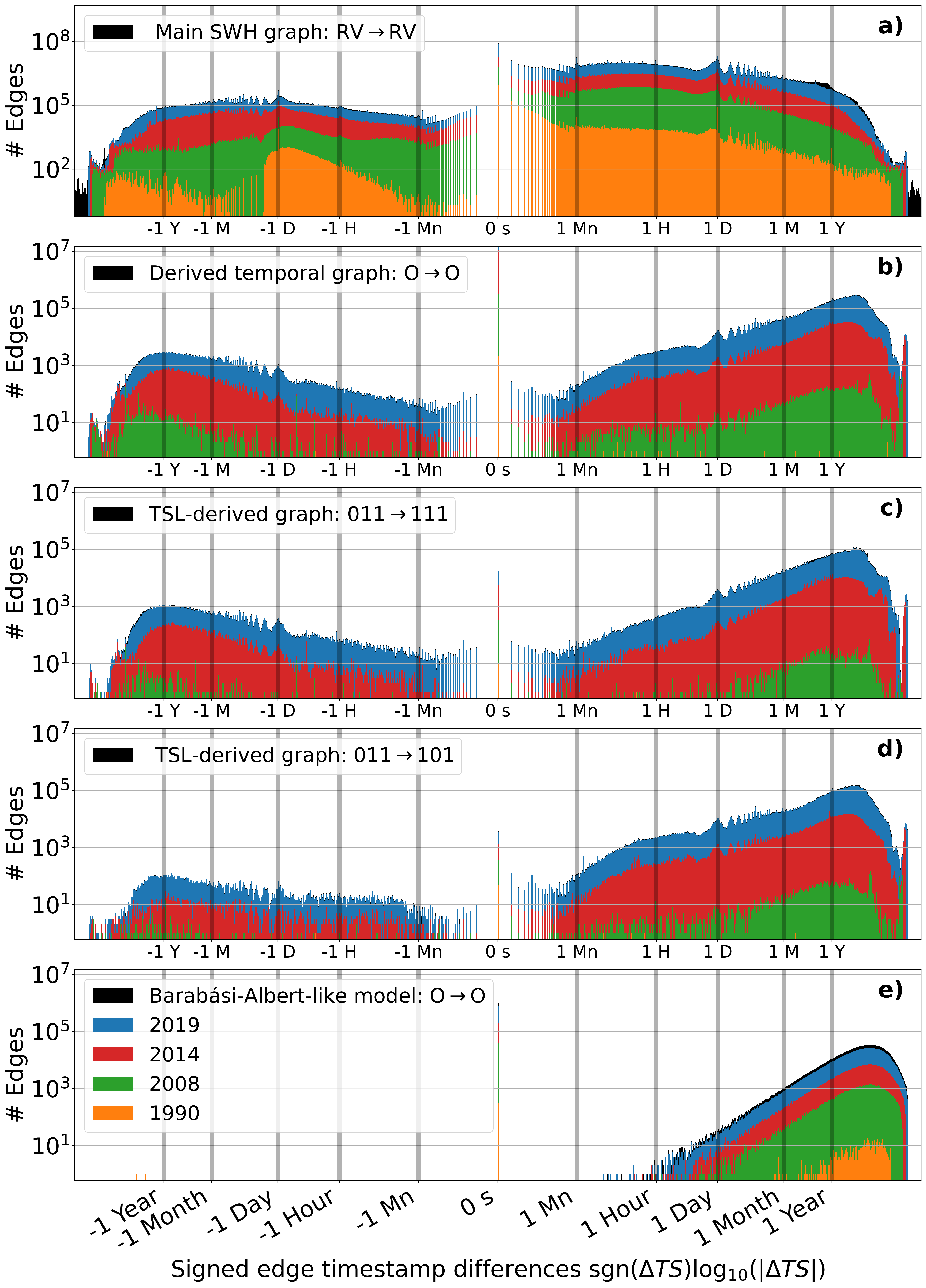}
    }
    }
    \caption{
Histograms of the signed edge timestamp differences ${\rm sgn}(\Delta TS)\log_{10}(|\Delta TS|)$.
From top to bottom, the panels correspond to: $RV{\to}RV$ edges of the {\it main graph};
$O{\to}O$ edges of the {\it temporal graph}; $011{\to}111$ and $111{\to}111$ edges from the
{\it TSL graph} after $TSL$ partitioning; and, for comparison, a modified Barabási--Albert
(Price) model (with two outgoing edges per new node, edges oriented according to node
appearance order, and timestamps defined to mimic the exponential growth of new nodes
observed in the {\it main graph}). The derived {\it temporal graph} used here has been
built using the {\it not-true-time} rule, which explains why histograms can exhibit a
non-zero probability for negative timestamp differences.
    }
    \label{fig:Comparison}
\end{figure}

The second characteristic discussed here concerns the dynamics of edge creation. Some
minimal models connect each new node only to preexisting nodes, while others also allow
the creation of new edges between already existing nodes upon the addition of each new
node, reflecting different structural growth mechanisms. In the empirical networks
studied here, an explicit edge creation timestamp is missing. However, since an edge can
only exist between two existing nodes, one can nevertheless infer certain features of the
underlying evolution rules by analyzing the histograms of the signed differences between
the appearance timestamps of source and target nodes for each edge, when such temporal
information exists. Figure~\ref{fig:Comparison} therefore shows histograms computed at
different times.

A first notable feature is the presence of edges with a negative signed difference, 
i.e., edges for which the source node appears after the target node. 
Such edges can only arise if the evolution rules allow this type of configuration.
In the modified Barabási--Albert (Price) model, this is not possible by construction, except for
the initial edges (see the three bins before $-1$ year in Fig.~\ref{fig:Comparison}, bottom
panel). By contrast, this is allowed by the {\it not-true-time} rule used here to build the
{\it temporal graph}.

The top panel, Fig.~\ref{fig:Comparison}a, corresponding to $RV{\to}RV$ edges of the
{\it main graph}, shows a density that decreases slowly up to approximately one year, and
then declines more sharply for larger values.\footnote{Note that the histograms are
constructed using fixed-width bins on a logarithmic scale, while the number of new nodes
grows exponentially.} An excess is also visible at one-day intervals and their multiples, which is
naturally related to the human activity underlying this network. In the top panel, the
dominant feature is the much greater weight of short time intervals below one day,
consistent with the fact that the creation of these nodes---and the edges between
them---is associated with daily software development activity.

The {\it temporal graph} and the {\it TSL graph} (middle three panels) exhibit the same
localized excess associated with daily periodicity, but display a different overall
picture, with a rapid increase in density. This highlights, as expected, the effect of
preferential attachment rules that favor the creation of edges pointing toward older
nodes---those that have had more time to accumulate incoming links from subsequently added
nodes. This behavior is also visible in the bottom panel (Fig.~\ref{fig:Comparison}e),
which shows the histogram obtained from the modified Barabási--Albert (Price) model and
exhibits a regular increase in density over the whole time range. 

A more detailed analysis
of the histogram of the {\it temporal graph} without $TSL$ partitioning
(Fig.~\ref{fig:Comparison}b) reveals several regimes corresponding to different time
scales---between 1 minute and 1 hour, 1 hour and 1 day, and beyond a few months---suggesting
the coexistence of distinct growth phenomena.
When distinguishing between the types defined by the $TSL$ partitioning, edges pointing
toward $101$ nodes (e.g., $011{\to}101$ edges in Fig.~\ref{fig:Comparison}d) exhibit a 
similar behavior. In contrast, Fig.~\ref{fig:Comparison}c shows a linear trend (on a
log--log scale) in the distribution of timestamp differences for $011{\to}111$ edges,
spanning time intervals from a few minutes to about one year, once the local excesses
associated with daily activity are excluded.

Taken together, these results show that the {\it temporal graph} derived from the {\it main graph}
captures a global growth pattern that is not visible in the subgraph of nodes with
native temporal attributes, and that is consistent with a minimal preferential-attachment
model. At the same time, the $TSL$ partitioning reveals that this apparent global behavior
may result from the superposition of distinct dynamical regimes, associated with
different classes of edges and time scales.

Even if the detailed growth rules still need to be investigated,
the presence or absence of {\it aging effects}, associated with a 
characteristic timescale associated for the loss of a node’s attractiveness, 
is an important feature. Such aging phenomena are known from 
the study of minimal models to be sufficient to prevent the persistence 
of scale-invariant properties at long times.
This effect is visible for the $RV$ nodes of the {\it main graph},
whose attractiveness decreases over time and drops even more sharply beyond approximately
one year (top panel, Fig.~\ref{fig:Comparison}).

For the $O$ nodes of the {\it temporal graph}, however, the situation is less clear: the
peak observed in the histograms, occurring for timestamp differences above five years,
remains comparable to the overall age of the network and therefore calls for a more
detailed analysis (see Discussion, Sec.~\ref{sec:discussion}).

\subsection{Scaling factor estimates}
\label{subsec:scalingfactor}
We now discuss the impact of the observed regime shifts, as well as of the proposed
partitionings, on the estimation of the scaling exponent associated with the ``tail'' of
the in-degree distribution, using one of the widely used methods~\cite{clauset_power-law_2009}.

This method is known to have many limitations~\cite{voitalov_scale-free_2019}. 
In particular, its first steps assume the existence of a scale-invariant regime characterized by
a distribution tail following a parametric power law. The scaling exponent is then
estimated by maximizing the likelihood function and using an ad hoc procedure to define a
threshold value associated with the beginning of the ``tail'' of the distribution. It can
therefore not be used to assess the scale-free hypothesis (see Discussion,
Sec.~\ref{sec:discussion}).  We focus here on the comparison of the estimates obtained with
the different graphs studied in this work, rather than on the method itself.

\begin{figure}
    {\centering
    \hbox{
    \includegraphics[width=1.0\linewidth]{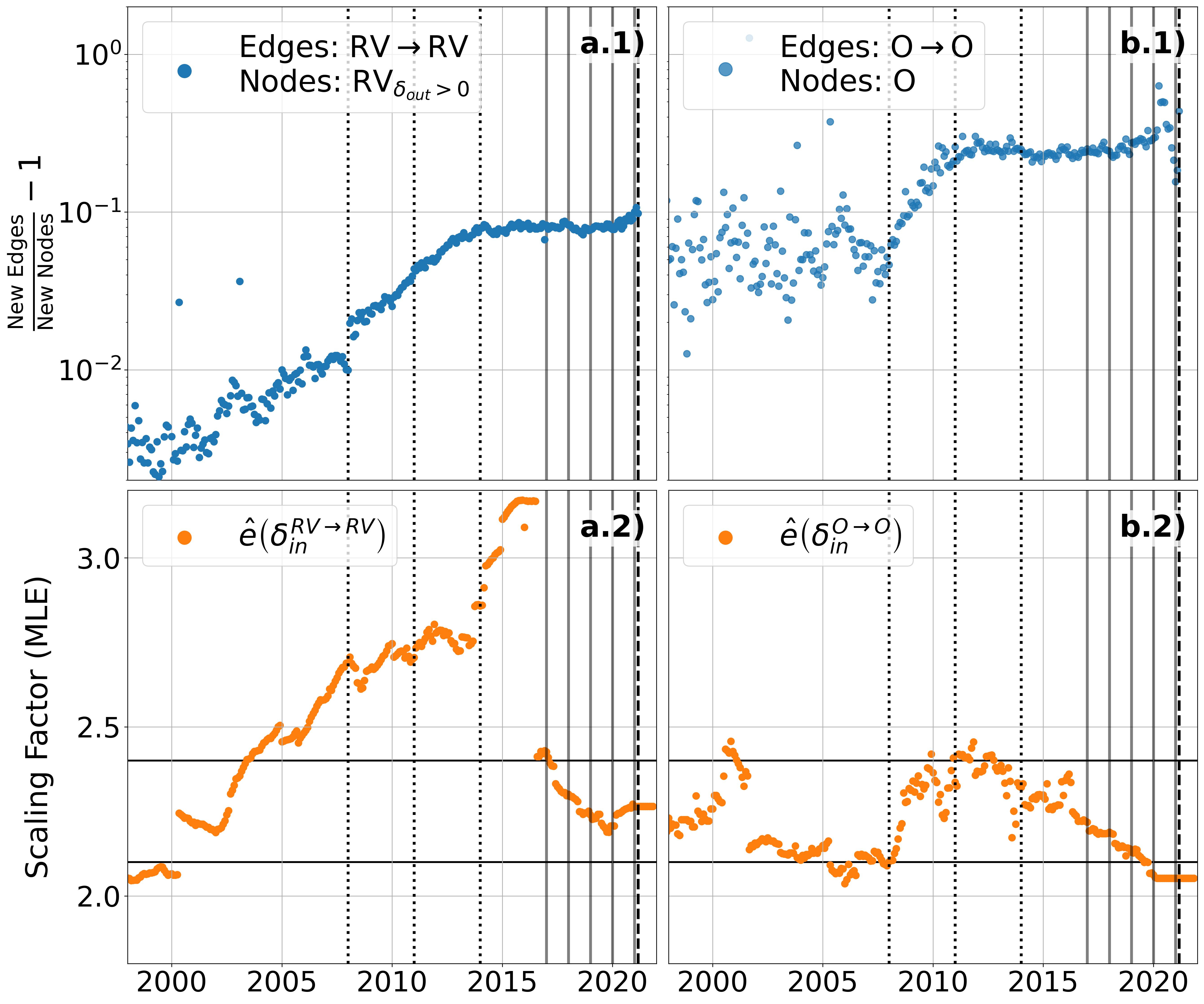}
    }
    }
    \caption{
(Left) Panel a.1 shows the ratio of new edges to new nodes over time for $RV$ nodes in the
$RV-RV$ subgraph of the {\it main graph}, highlighting changes in growth regimes
occurring in 2008, around 2011, and from 2014 onward (dotted lines). Panel a.2 shows the
estimated power-law exponent $\hat{e}(\delta_{in})$, computed using steps 1 and 2 described
by Clauset et al.~\cite{clauset_power-law_2009}, under the assumption that the distribution
tail follows a parametric form $df(\delta) \propto \delta^{-e}$. 
(Right) Same representations (panels b.1 and b.2) for $O$ nodes in the {\it temporal graph}.
The degree distributions in this network appear more regular and less affected by
outliers, yielding a more robust estimate of the scaling exponent, although this estimate
should not be interpreted as evidence for a scale-free regime.
    }
    \label{fig:DegreeOOmerged_bottom}
\end{figure}

Fig.~\ref{fig:DegreeOOmerged_bottom} displays the scaling exponents estimated over time for
the in-degree distribution of $RV$ nodes in the $RV-RV$ subgraph of the {\it main
graph} (panels a.1 and a.2), as well as for $O$ nodes in the derived {\it temporal graph}
(panels b.1 and b.2). Due to the presence of outliers in the distribution
associated with $RV$ nodes (Fig.~\ref{fig:DegreeOOmerged_top}a), the estimation method
appears significantly more sensitive, exhibiting strong temporal fluctuations (see
Fig.~\ref{fig:DegreeOOmerged_bottom}, panel a.2, end of 2016). This makes any association
with the growth dynamics observed in panel a.1 almost impossible. The sensitivity of the
method proposed by Clauset et al.\ is discussed in more detail in Supplementary
Material~8.

As previously mentioned, the degree distributions associated with the derived {\it
temporal graph} and the {\it TSL graph} (Fig.~\ref{fig:DegreeOOmerged_top}) exhibit greater
regularity. The scaling exponents estimated over time for the {\it temporal graph} vary
more smoothly (Fig.~\ref{fig:DegreeOOmerged_bottom}, panel b.2) and exhibit an increase
that aligns with the observed transition in the number of new edges per new node (same
figure, panel b.1, between 2008 and 2011). Both exponent estimates (panels a.2 and b.2)
decrease when approaching the dataset export date. This leaves open the possibility that the
anomalies identified in the dataset and discussed previously (Fig.~\ref{fig:ratioRVRV},
end of 2017 and about 15 months before export), or a more general relaxation process
characterizing the regime shifts, may influence these measurements. A more detailed
analysis, including the evaluation of scaling exponents for the derived graphs after $TSL$
partitioning, is provided in Supplementary Material~11.

With a few exceptions, existing studies based on minimal models focus on long-term or
steady-state regimes and therefore provide limited insight into the nature of regime
shifts and, more broadly, raise the question of how such shifts—or anomalies, which are in
fact common in real-world networks—affect the conditions under which network properties
can be observed.

\section{Application to the APS citation dataset}
\label{sec:application}
Before concluding, we briefly discuss the generality of our findings and the relevance of
this study for the development of a generic methodology to analyze real-world growing
networks and compare them with minimal models. To this end, we apply the same approach
used for the Software Heritage dataset to a different empirical system, namely the APS
citation network.

The {\it APS Data Sets for Research}\footnote{\url{https://journals.aps.org/datasets}} is
dataset made available upon request by the American Physical Society. It spans
approximately 130 years of bibliographic metadata for APS journal articles, together with
the corresponding citation links. References to articles published outside the APS corpus
are not available in the dataset. The resulting APS citation network shares several similarities with the
{\it main graph} analyzed in this study, in particular a relatively simple growth
dynamics: nodes and edges are created once and for all, and new directed edges typically
connect newly introduced nodes to preexisting ones, although some exceptions may occur.

We started from the 2022 export of the APS dataset, which includes articles and citations
up to the end of 2022. It contains nearly 725,157 publications, of which 720,234 have a
valid timestamp, and 9,758,055 associated citations between articles published in APS
journals. In the scope of this study, we performed a straightforward import of the data,
without, for instance, distinguishing between publications from different journals or
including author information (each author could be represented as a node of type
``Author''), even though this would have made sense in the context of a more detailed
study.

The APS dataset has been the subject of numerous investigations~\cite{price_general_1976,
krapivsky_organization_2001,redner_citation_2005,sheridan_preferential_2018}. These works
examine either the full APS dataset or subsets of it, and more broadly explore the role of
preferential attachment and cumulative advantage mechanisms in the structure and evolution
of citation networks. The underlying growth mechanisms support strong assumptions about
the presence of {\it aging effects}, which under certain conditions, can lead to 
non-scale-free in-degree distributions. For instance, {\it
Supplementary Note 3} of Sheridan et al.\ (2018)~\cite{sheridan_preferential_2018} provides a
formal proof that incorporating aging into the preferential attachment 
growth model ($P \propto k/\ln k$ for large
degrees) leads to an in-degree distribution that asymptotically follows a log-normal law
for large degrees. However, this result relies on a key assumption: {\it ``The mean value
$m$ of the $m_t$'s is constant over time with finite variance as $t$ becomes larger.''} We
will not delve into the implications of this assumption here, nor discuss in detail its
consequences for the analysis of this dataset—a topic we leave for future
work. Nevertheless, we emphasize that this assumption is representative of commonly
accepted hypotheses when comparing the structural properties of real-world citation
networks with minimal growth models.

\begin{figure}
    {\centering
    \hbox{
    \includegraphics[width=1.0\linewidth]{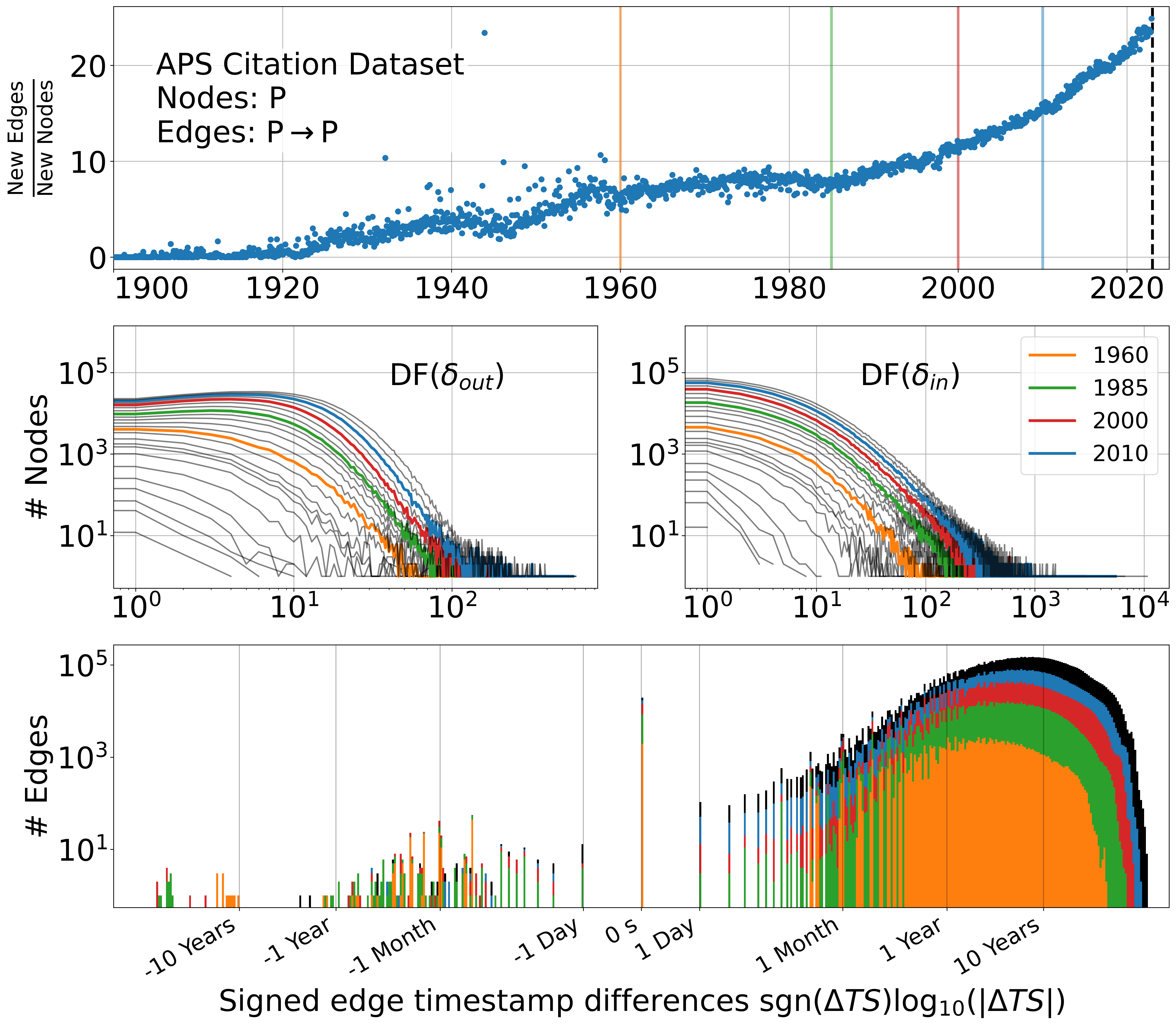}
    }
    }
    \caption{
This figure reproduces the main representations discussed earlier, applied to the APS
citation dataset (2022 export). (Top) Average number of new edges per new node per month
between 1900 and 2022. Vertical lines indicate specific dates of interest (1960, 1985,
2000, 2010) discussed here and in Supplementary Material Section~13 dedicated to this
dataset. A clear exponential increase in the average number of edges per node is observed
starting around 1985. (Mddle) Cumulative out-degree (left) and in-degree (right)
distributions over time. The same color code is used to highlight the key years. The
evolution of the out-degree distribution is particularly insightful, as it reveals a
change in the characteristics of the underlying instantaneous distribution, and therefore
a shift in the growth dynamics of the network. (Bottom) Histograms of the time differences
between source and target node timestamps over time, confirming the near-total absence of
edges originating from preexisting nodes. While the dataset stores timestamps at the
resolution of one second, the actual minimum meaningful difference is one day (or zero for
same-day citations). For readability, the histograms are centered by normalizing the timestamp
difference ($\Delta TS$) by a constant (chosen here as $1/5$ day), effectively shifting
them towards the center. The histograms accumulate over time, with the same color code
used to distinguish the key years.
    }
    \label{fig:APS}
\vspace{-0.5cm}
\end{figure}

To replicate our analysis in a more generic setting, we first construct a synthetic
summary representation based on the main observables discussed above. Figure~\ref{fig:APS}
shows a clear regime change around 1985 in the ratio of new edges to new nodes: nearly
constant before this date (1960--1980), it subsequently exhibits an approximately
exponential growth. This transition coincides with a change in the shape of the out-degree
distribution: the time-aggregated degree distributions computed from the origin of the
network up to different times no longer exhibit the same shape in log--log representation
for all degrees, thereby ruling out the stationarity assumption.

Possible explanations include changes in citation practices—potentially linked to the
increasing role of bibliometric indicators~\cite{garfield_new_1963} 
such as impact factors and international
rankings—as well as structural biases in the dataset, which records only citations between
APS articles. Both factors could contribute to an apparent increase in the internal
citation rate independent of genuine structural change in the scientific literature.

\begin{figure}
    {\centering
    \hbox{
    \includegraphics[width=1.0\linewidth]{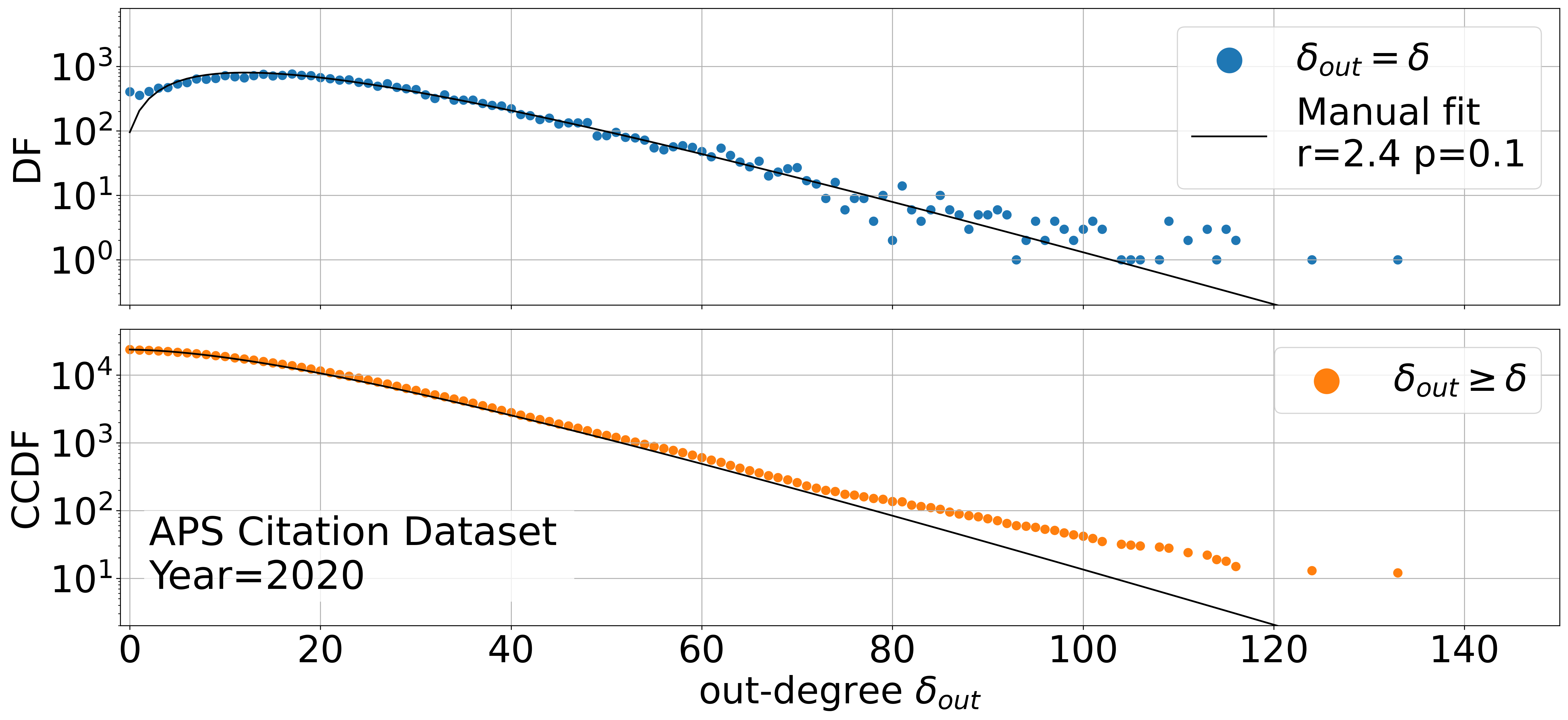}
    }
    }
    \caption{
Out-degree distribution (Top) and associated CCDF (Bottom) extracted from the APS citation
dataset for articles published in 2020. The solid line corresponds to a negative binomial
distribution with parameters $r = 2.4$ and $p = 0.1$, which were manually fitted to match
the central part of the out-degree distribution. One can observe an excess of zero
out-degree nodes, as well as an overrepresentation of very high out-degree nodes.
    }
    \label{fig:APSdout2020}
\vspace{-0.5cm}
\end{figure}

The topological partitioning of the APS network reveals the existence of multiple
components in the out-degree distribution, with the largest component well described by a
negative binomial distribution. The out-degree distribution restricted to articles
published in 2020 (Fig.~\ref{fig:APSdout2020}) exhibits two notable anomalies: an excess of
zero out-degree nodes, and an overrepresentation of very high out-degree nodes, partly
attributable to specific journals (e.g., {\it Reviews of Modern Physics}, see Supplementary
Material~13).

Together, these observations make the APS dataset a second empirical example in this
study, underscoring the importance of characterizing temporal variations in degree growth
regimes—and the possible coexistence of distinct generative mechanisms—before comparing
empirical networks with minimal theoretical models. Any comparison involving the APS
citation dataset must account for both the regime change around 1985 and the richer
initial conditions of the network, which had already experienced substantial evolution
since at least 1965 under different growth dynamics.

\section{Discussion}
\label{sec:discussion}
In this study, we analyzed the growth properties of the SWH dataset 
under minimal assumptions regarding the underlying growth rules. This dataset forms 
a very large network including billions of nodes and spanning several decades of 
public source code development history.

Previous \textit{static} studies of connected component size distributions and degree
distributions in this system revealed heavy-tailed in-degree distributions and non-trivial
multi-scale aggregation processes, supporting the emergence of structures of all sizes and,
possibly, the existence of a scale-free regime. However, expected aging effects—such as
technological obsolescence—challenge the assumptions underlying linear preferential
attachment models, which are known to be among the key conditions for the emergence of
scale-free behavior.

Addressing this issue therefore requires a non-static analysis, which is made difficult by
the fact that some nodes in this heterogeneous network do not have temporal attributes, and
raises the question of the coexistence or competition of multiscale growth mechanisms.

The analysis of subgraphs of the {\it main graph} and of the derived {\it temporal} and {\it TSL graphs},
based on topological and temporal partitionings, made it possible to capture several key
aspects of the growth rules at play in this network. It highlights changes
in the growth regime, especially following the widespread adoption of distributed version
control systems (notably Git); disentangles long-term attachment mechanisms, revealed by the
histograms of signed edge timestamp differences in the {\it temporal graph}, from short-term
attachment time scales characterizing edges between $RV$ nodes; and shows the sensitivity
of widely used scaling-exponent estimation methods to the presence of numerous outliers and
to regime changes in the observed distributions, while also indicating that the proposed
partitioning helps mitigate these effects.

Several limitations may affect the interpretation of the results. The chosen temporal
partitioning strategy is not unique and may introduce biases. 
It is non-causal, as it evaluates origin node sizes based on the number of
reachable $RV$ nodes at a time close to the dataset’s extraction date. As a result, future
exports—including additional $RV$, $RL$, and $O$ nodes—may yield {\it temporal graphs} in
which edges obtained from a previous export are not guaranteed to persist. Moreover, the
origin size distributions used in the partitioning exhibit non-trivial, possibly
heavy-tailed behavior from the outset. Thus, the robustness of the observed topological
properties and inferred evolution rules must be further challenged by verifying their
consistency across alternative partitioning strategies—and ideally, under causal temporal
partitioning, before being interpreted as intrinsic properties of the system.

Another limitation of the temporal partitioning used here is that it may partially mask
aging effects by favoring forked projects (e.g., LibreOffice over OpenOffice), which
typically have more incoming edges and larger current sizes than their original
counterparts. Moreover, even if long-term attachment mechanisms do exist, their effect
would still need to be demonstrated through a statistically significant correlation with
the in-degree distribution, for instance by inferring the underlying growth rules, or by
linking the properties of edge timestamp differences and degree distributions through a
theoretical study of minimal models. This point is left for future work, together with a
more systematic investigation of the influence of different partitioning rules, in
particular regarding the treatment of inheritance and the use of the arrow of time to
orient edges, even though the elements available in the supplementary materials suggest
that the observed differences remain limited within the scope of this study.

We then applied the same approach developed for the SWH dataset to the APS citation network.
This analysis reveals that, contrary to common assumptions, the APS dataset exhibits a
significant change in its evolution rules before and after 1985. The most recent regime is
characterized by an accelerated growth in the number of new edges per new node. Although
such a regime is not sustainable in the long term and suggests a further transition either
already ongoing or yet to come, it highlights some of the methodological limitations
encountered when relying on previous studies of minimal growth models.

Indeed, several models with accelerated growth have been proposed (see Sec.~VIII.B
in~\cite{albert_statistical_2002}), and their theoretical and numerical analyses confirm
that they can still lead to the emergence of scale-invariant regimes. These results 
demonstrate that the mere existence of an accelerated growth regime
is not a sufficient condition to suppress the emergence of scale-free behavior in the
system. However, it should be noted that, to induce accelerated growth, these
models typically rely on the creation of edges after the appearance of the nodes they
connect, and are therefore potentially non-causal in the sense defined above, unless
reliable timestamps for edge creation are available—which is frequently not the case in
empirical datasets.

In a similar vein, long-term memory effects associated with initial conditions may
critically influence the observability of the expected asymptotic regime(s).

\section{Conclusion}
\label{sec:conclusion}

In this work, we proposed and applied temporal and topological partitioning strategies
to confront large empirical growing networks with minimal growth models, using the Software
Heritage and APS datasets as case studies.

The evidence presented here shows that heterogeneity, partial temporal information, and
non-stationary effects strongly challenge naive or purely static comparisons with minimal
models and call for a careful interpretation of standard observables such as degree
distributions or scaling exponents.

More generally, this study highlights the need for refined analysis tools and genuinely
causal modeling frameworks to properly account for transient regimes, structural
transitions, and the interplay of multiple growth mechanisms in real-world evolving
networks.

\clearpage
\section{Appendix: Algorithms and list of acronyms}
\label{sec:algorithms}
\subsection{Algorithms}
We present here the different algorithms used to derive the {\it temporal graph} from the {\it main graph}..

\begin{algorithm}[h!]
\caption{\textsc{OriginSizes}: Compute the size of each origin}
\KwIn{Graph $G=(V,E)$ with origins $O$ and timestamped nodes $RV \cup RL$}
\KwOut{Size $S(o)$ for all origins $o \in O$}

\ForEach{$o \in O$}{
    $S(o) \gets 0$\;
    $Visited \gets \emptyset$\;

    \ForEach{$v$ reachable from $o$ in $G$ via directed paths}{
        \If{$v \notin Visited$}{
            $Visited \gets Visited \cup \{v\}$\;
            \If{$type(v)\in\{RV,RL\}$}{
                $S(o) \gets S(o)+1$\;
            }
        }
    }
}
\Return{$S$}\;
\end{algorithm}

\begin{algorithm}[h]
\caption{\textsc{PartitionRVRL}: Partition of $RV$ and $RL$ nodes by decreasing origin size}
\KwIn{Graph $G$, origin sizes $S(o)$}
\KwOut{Assignment $origin(v)$ for all $v \in RV \cup RL$}

\ForEach{$v \in RV \cup RL$}{
    $A(v) \gets$ all origins reaching $v$\;
    Select $o \in A(v)$ with maximal $S(o)$\;
    $origin(v) \gets o$\;
}
\Return{ $origin$ }\;
\end{algorithm}

\begin{algorithm}[h!]
\caption{\textsc{OriginTemporalities}: Assign temporal attributes to $O$ nodes}
\KwIn{Assignments $origin(v)$, timestamps $t(v)$ for $v\in RV\cup RL$}
\KwOut{Temporal values $\hat{t}(o)$ for each origin $o$}

\ForEach{$o \in O$}{
    $T(o) \gets \{\, t(v)\mid v\in RV\cup RL,\ origin(v)=o\,\}$\;
    \eIf{$T(o)$ is empty}{
        $\hat{t}(o) \gets +\infty$\;
    }{
        $\hat{t}(o) \gets \min T(o)$\;
    }
}
\Return{$\hat{t}$}\;
\end{algorithm}

\begin{algorithm}[h]
\caption{\textsc{DerivedGraphNoTT}: 
Build the derived origin graph with ($I$) or without ($WI$) the inheritance rule.
In the WI variant, an edge $o_i \to o_j$ is added whenever at least 
one explicit $RV/RL$ edge connects a node assigned to $o_i$ to a node 
assigned to $o_j$ (including $i=j$).
In the I variant, an edge $o_i \to o_j$ is added whenever a directed path exists 
from $o_i$ to $o_j$.
Edges are unweighted and are added only if not already present. 
The output derived graph does not use the true-time rule.}

\KwIn{Graph $G$, assignments $origin(v)$, mode $\in \{\mathrm{WI},\mathrm{I}\}$}
\KwOut{Derived graph $G^{\mathrm{mode,NoTT}}$}

\If{$mode = \mathrm{WI}$}{
    Initialize $G^{\mathrm{WI,NoTT}}$ with node set $O$\;
    \ForEach{edge $(u \to v)$ in $G$ with $u,v \in RV \cup RL$}{
        $o_u \gets origin(u)$\;
        $o_v \gets origin(v)$\;
        \If{edge $(o_u \to o_v)$ does not already exist in $G^{\mathrm{WI,NoTT}}$}{
            add edge $(o_u \to o_v)$ to $G^{\mathrm{WI,NoTT}}$\;
        }
    }
    \Return{$G^{\mathrm{WI,NoTT}}$}\;
}

\If{$mode = \mathrm{I}$}{
    Initialize $G^{\mathrm{I,NoTT}}$ with node set $O$\;
    \ForEach{$o_i \in O$}{
        \ForEach{$v$ reachable from $o_i$ in $G$ with $v \in RV \cup RL$}{
            $o_j \gets origin(v)$\;
            \If{edge $(o_i \to o_j)$ does not already exist in $G^{\mathrm{I,NoTT}}$}{
                add edge $(o_i \to o_j)$ to $G^{\mathrm{I,NoTT}}$\;
            }
        }
    }
    \Return{$G^{\mathrm{I,NoTT}}$}\;
}

\end{algorithm}

\begin{algorithm}[h]
\caption{\textsc{DerivedGraphTT}: Build the true-time version of a derived origin graph $G^{\mathrm{WI,NoTT}}$ or
$G^{\mathrm{I,NoTT}}$. All edges are kept, but their direction is reoriented to follow the true-time ordering
$\hat{t}(o)$ of origins.}

\KwIn{mode $\in \{\mathrm{WI},\mathrm{I}\}$, derived graph $G^{\mathrm{mode,NoTT}}=(O,E^{\mathrm{mode,NoTT}})$, temporal
attributes $\hat{t}(o)$}
\KwOut{True-time graph $G^\mathrm{mode,TT}$}

Initialize $G^{\mathrm{mode,TT}}$ with node set $O$\;

\ForEach{edge $(o_i \to o_j)$ in $E^{\mathrm{mode,NoTT}}$}{
    \If{$\hat{t}(o_i) < \hat{t}(o_j)$}{
        add edge $(o_i \to o_j)$ to $G^{\mathrm{mode,TT}}$\;
    }
    \Else{
        add edge $(o_j \to o_i)$ to $G^{\mathrm{mode,TT}}$\;
    }
}
\Return{$G^{\mathrm{mode,TT}}$}\;
\end{algorithm}

\begin{algorithm}[h]
\caption{\textsc{DerivedGraph}: Global construction of derived origin graphs. 
Given inheritance mode $modeI \in \{\mathrm{WI},\mathrm{I}\}$ and temporal mode 
$modeT \in \{\mathrm{NoTT},\mathrm{TT}\}$, the algorithm outputs 
$G^{modeI,modeT}$.}

\KwIn{Graph $G$, inheritance mode $modeI$, temporal mode $modeT$}
\KwOut{Derived origin graph $G^{modeI,modeT}$}

$S \gets$ OriginSizes$(G)$\;                       
$origin \gets$ PartitionRV\_RL$(G,S)$\;      
$\hat{t} \gets$ OriginTemporalities$(origin,G)$\;   
$G^{modeI,NoTT} \gets$ DerivedGraphNoTT$(G,origin,modeI)$\;   

\If{$modeT = \mathrm{NoTT}$}{
    \Return{$G^{modeI,NoTT}$}\;
}

$G^{modeI,TT} \gets$ DerivedGraphTT$(G^{modeI,NoTT},\hat{t})$\;
\Return{$G^{modeI,TT}$}\;

\end{algorithm}

\subsection{List of Acronyms}

\begin{tabular}{ll}
\textbf{Acronym} & \textbf{Meaning} \\
\hline
SWH & Software Heritage \\
APS & American Physical Society \\
BA & Barabási--Albert \\
RV & Revision node (SWH dataset)\\
RL & Release node (SWH dataset)\\
O & Origin node (SWH dataset)\\
P & Publication node (APS citation dataset)\\
DAG & Directed Acyclic Graph \\
CCDF & Complementary Cumulative Distribution Function \\
DF & Distribution Function \\
$\delta_{in}$ & in-degree \\
$\delta_{out}$ & out-degree \\
TT & True-Time \\
NoTT & Not-True-Time \\
I & With inheritance \\
WI & Without inheritance \\
O:TSL($\delta_m$) & Topological classification based on (T, S, L,$\delta_m$) \\
TSL & Short acronym similar to  O:TSL($\delta_m$) \\
$\delta_m$ & Depth parameter in the $TSL$ classification\\
T & In-degree criteria (Target) in the $TSL$ classification\\
S & Out-degree criteria (Source) in the $TSL$ classification\\
L & Self-loop criteria (Loop) in the $TSL$ classification\\
\end{tabular}

See Table~\ref{tab:notations} for the notations and definitions used for the {\it main graph}, the {\it temporal graph}, and the {\it TSL graph}.

\clearpage
\bibliographystyle{plain}
\bibliography{main}
\end{document}